\documentstyle[aps,twocolumn,epsfig]{revtex}

\begin{document}
\draft

\title {Scaling Properties of Fluctuations in\\
Human Electroencephalogram}

\author{Rudolph C. Hwa}
\address{Institute of Theoretical Science and Department of
Physics\\ University of Oregon, Eugene, OR 97403-5203, USA}

\author{Thomas C. Ferree\cite{byline}}

\address{Electrical Geodesics, Inc., Riverfront Research Park,
Eugene, OR 97403}

\date{\today}
\maketitle

\begin{abstract} The fluctuation properties of the human
electroencephalogram (EEG) time series are studied using
detrended fluctuation analysis.  For all 128 channels in each of
18 subjects studied, it is found that the standard deviation of
the fluctuations exhibits scaling behaviors in two regions.
Topographical plots of the scaling exponents reveal the spatial
structure of the nonlinear electrical activities recorded on the
scalp.  Moment analyses  are performed to extract the gross features
of all the  scaling exponents. The correlation between the two
scaling exponents in each channel is also examined. It is found that
two indices can  characterize the overall properties of the
fluctuation behaviors of the brain dynamics for every subject and
that they vary widely across the subjects.

\end{abstract}


\section{Introduction}

The scalp electroencephalogram (EEG) provides a wealth of
information about human brain dynamics.  The complex nature
of brain dynamics results in a high degree of fluctuations in
both the spatial and temporal aspects of the EEG signals. To
extract the salient properties from the data is the primary
objective of any method of analysis.  We present in this paper
a novel method that explores the scaling behavior of the
fluctuations and uses moment analysis to reduce the complexity
of the results obtained.

The most common methods of EEG time series analyses are
event-related time ensemble averaging and Fourier decomposition,
both of which are based implicitly on assumptions of
linearity \cite{eeg,nunez}.  Since the physiological
mechanisms underlying the scalp EEG are generally nonlinear,
they can generate fluctuations that are not best described
by linear decomposition.  Moreover, the resting EEG always
displays a broad-banded power spectrum, so in Fourier
analysis one must arbitrarily define frequency bands
($\delta, \theta, \alpha, \cdots$) which may not actually
delineate different dynamical mechanisms.  Wavelet analyses
have also been applied to examine EEG time series \cite{bla},
but at a sacrifice of the ability to describe long-range
temporal correlations.  Chaos analyses have been applied
to quantify the nonlinear behavior of the brain
\cite{chaos,jan,leh}, but typically require a long period
of time to compute attractor properties for a single time series.
Moreover, chaos-based approaches assume the existence of
low-dimensional attractors, and this is probably not generally a
valid  assumption for the brain.  In this paper, we discuss a
method that analyzes the fluctuations in EEG over a short
period of time (around 10s), and makes use of the information
conveyed by all 128 channels on the scalp.  We show the
existence of scaling behaviors of certain measures of
fluctuations in all channels and in all subjects.   We
propose two global measures of the spatio-temporal signals
that characterize the distinctive nature of EEG for each subject.

The study of scaling behavior emphasizes the relationship
across time scales.  We aim to find what is universal among
all channels as well as what varies among them.  The former
is obviously important by virtue of its universality for a
given subject; how that universal quantity varies from
subject to subject is clearly interesting.  What varies
from channel to channel is perhaps even more interesting,
since it has implications for describing focal features
which could have functional or clinical relevance.

Our procedure is to focus initially on one channel at a time.
Thus it is a study of the local temporal behavior and the
determination of a few parameters (scaling exponents) that
effectively summarize the fluctuation properties of the time
series. The second phase of our procedure is to describe the
global behavior of all channels and to arrive at two numbers that
summarize the variability of these temporal measures across the
entire scalp surface.  This dramatic data reduction  necessarily
trades detail for succinctness, but such reduction  is exactly
what is needed to allow easy discrimination between  brain
states.

The emphasis in this paper is on the method of analysis. It is not
our aim here to perform clinical and cognitive analyses. Due to
the fact that the EEG data available to us are short in time
duration and few in the number of subjects studied, it is not
feasible for us to make reliable inference on the physiological
implications of our findings. Nevertheless, the data are
sufficient for the extraction of interesting behaviors that are
channel dependent as well as subject dependent.

\section{Detrended Fluctuation Analysis}

The specific method we use in the first phase is detrended
fluctuation analysis (DFA). This analysis is not new. It was
proposed for the investigation of correlation properties in
non-stationary time series and applied to the studies of
heartbeat \cite{peng} and DNA nucleotides \cite{pbh}.  It has
also been applied to EEG \cite{wat}, but with somewhat different
emphases than those presented here.  Since the  analysis
considers only the fluctuations from the local  linear trends,
it is insensitive to spurious correlations  introduced by
non-stationary external trends.  By examining  the scaling
behavior one can learn about the nature of  short-range and
long-range correlations.

Let an EEG time series be denoted by $y(t)$, where $t$ is
discrete time ranging from 1 to $T$. Divide the entire range of
$t$ to be investigated into $B$ equal bins, discarding any
remainder, so that each bin has $k={\rm floor}(T/B)$ time points.
Within each bin, labeled $b\ (b=1,\cdots, B)$, perform a least-square
fit of $y(t)$ by a straight line, $\overline y_b(t)$, i.e.,
$\overline  y_b(t)=$ Linear-fit$[y(t)]$ for $(b-1)k<t\leq bk$.
That is the semi-local trend for the $\it b$th bin. Define
$F_b^2(k)$ to be the variance of the fluctuation $y(t)$ from
$\overline y_b(t)$ in the $b$th bin, i.e.,
\begin{equation}
F_b^2(k) = {1\over k} \sum_{t=(b-1)k+1}^{bk} [y(t)-\overline
y_b(t)]^2
\label{(1)}
\end{equation}
It is a measure of the semi-locally detrended fluctuation in bin $b$.
The average of $F_b^2(k)$ over all bins is
\begin{equation}
 F^2(k)={1\over B}\sum_{b=1}^B\,F_b^2(k).
\label{(2)}
\end{equation}
$F(k)$ is then the RMS fluctuation from the semi-local trends in
$B$ bins each having $k$ time points.

The study of the dependence of $F(k)$ on the bin size $k$ is the
essence of DFA \cite{peng,pbh}. If it is a power-law behavior
\begin{equation}
 F(k) \propto k^{\alpha} ,  \label{(3)}
\end{equation}
then the scaling exponent $\alpha$ is an
indicator of the correlations of the fluctuations in EEG,  which
depends on the relationship of these fluctuations across time
scales. Since DFA considers only the fluctuations from
the semi-local linear trends, it is insensitive to spurious correlations
introduced by non-stationary external trends.  This is a practical
advantage since EEG acquisition systems often suffer from slow
drifts associated with gradual changes in the quality of electrode
contact to the skin.  The analysis also liberates our result
from the dependence on the overall magnitude of the voltage
$y(t)$ recorded by each probe, which is an advantage since
overall signal amplitude can vary across subjects,
presumably due to differences in skull conductivity and other factors.

Resting EEG data were collected for 18 subjects using a
128-channel commercial EEG system, with scalp-electrode
impedences ranging from 10 to 40 k$\Omega$. The data were
hardware filtered between 0.1 and 100 Hz, then digitized at
250 points/sec. After acquisition, $T\approx 10 $s lengths of
simultaneous time series in all channels are chosen, free of
artifacts such as eye blink and head movements.  At each time
point, the average across all electrodes was subtracted, to
remove approximately the effect of the reference electrode
\cite{nunez}.  We investigate the range of $k$
from 3 to 500 in approximately equal steps  of $\ln k$.

In Fig.\,1 we show three typical time series $y(t)$ in three
widely separated channels for subject A, labeled 1-3, for
brevity.  While it is clear that both channels 2 and 3
have substantial 10 Hz  oscillations after 0.2s, it is much
less apparent that there exist any scaling behaviors in all three
channels.  The corresponding values of $F(k)$ are shown in  the
log-log plot in Fig.\,2.  Evidently, the striking feature  is
that there are two scaling regions with a discernible  bend when
the two slopes in the two regions are distinctly  different.
With rare exceptions this feature is found in  all channels for
all subjects.  Admittedly, the extents of  the scaling regions
are not wide, so the behavior does not  meet the qualification
for scaling in large critical systems  or in fractal geometrical
objects. However, since the behavior  is so universal across
channels and subjects, and since the temporal scales involved are
physiologically relevant,  the scaling behavior is a feature  of
EEG that conveys an important property of the brain activity  and
should not be overlooked.

\section {Scaling and Nonscaling Properties}

The fact that there exist two scaling regions suggests that the
lack of scaling in the region between the two implies the
existence of some significant time scale. From Fig.\ 1 one indeed
sees roughly periodic oscillations in Ch.\ 2 and 3. One may
therefore be tempted to think that if, instead of considering the
fluctuations from the linear semi-local trends $\bar y_b(t)$, one
studies the fluctuations from periodic oscillations, then the
bend would disappear and the two scaling regions might be joined
to become one. However, even if that were true, such a procedure
should not be used for two reasons. First, not all channels
exhibit obvious oscillatory behaviors with definite frequencies.
Channel 1 in Fig.\ 1 is one such example. Whatever detrending one
chooses should be universally applied to all channels in order to
avoid introducing discrepencies across the channels due to
external intervention. Second, to determine the frequency of the
oscillatory trend requires a Fourier analysis, which is precisely
what our approach attempts to circumvent. To decide on a sinusoidal
wave of a particular frequency as reference for detrending
involves arbitrariness and is unlikely to lead to any simplification
in the global picture. The simplest and unbiased approach is to use
the semi-local linear trends, as we have done.

To quantify the scaling behavior, we perform a linear fit in
Region I for $1<{\ln}k < 2.5$ and denote the slope by
$\alpha_1$, and similarly in Region II for $3.5 < {\ln}k < 5.75$ with
slope denoted by $\alpha_2$.  Visual inspection for each of the
18 subjects verifies that fitting this way does a remarkably
good job of characterizing the slopes in the two regions.
Knowing the two straight  lines in each channel allows us to
determine the location of  their intercept, $\ln\kappa$, which
gives a good approximation for the position  of the bend in
${\ln}k$. We find that, whereas $\alpha_1$  and $\alpha_2$ can
fluctuate widely from channel to channel, $\kappa$  is limited to
a narrow range in most subjects. The average value of ln$\kappa$
for each subject ranges from 2.6 to 3.6, with a grand average
across subjects to be approximately 3.1. It should, however, be
noted that when $\alpha_1$ and $\alpha_2$ are nearly the same, as
is the case for Ch.\ 1 in Fig.\ 2, the determination of $\kappa$
by the intersection of the two straight lines is not reliable.
Nevertheless, it is visually clear that the bend occurs in the
vicinity of ln$\kappa = 3.1$.

Since scaling behavior means that the system examined has no
intrinsic scale, scale noninvariance at $\kappa$ implies that
$\kappa$ is related to a dominant frequency of oscillation in the
time series. It is at this point that a contact can be made with
the usual Fourier analysis. Although our analysis focuses on scale
invariant quantities, i.e., the dimensionless scaling exponents, it
is worth digressing momentarily to establish this contact.  To do
this, we loosely associate the time scale $\kappa$ with the period
of a sine wave with frequency $f$. If the data acquisition rate is
denoted by $r$, then the frequency $f$ corresponding to  $\kappa$
is
\begin{equation}
f=r/\kappa.		  \label{(4)}
\end{equation}
For our data acquisition we have $r=250$ points/sec.
For the across-subject average of ln$\kappa = 3.1$, we get from
Eq.\ (4) $f=11.3$ Hz. That is in the middle of the traditional
$\alpha$ (8-13 Hz) EEG frequency band.  Thus the dominant periodic
oscillation apparent in Fig.\ 1 does reveal itself in the study of
the scaling behavior.  If one's interest is in the frequency
content of the EEG time series, then Fourier analysis is more
direct. However, if the interest is in the fluctuation properties
and their relationships across time scales, then DFA is
more effective.   Hereafter, frequency will play no essential role
in the remainder of this paper.

For each subject we have 128 pairs of values of $(\alpha_1,
\alpha_2)$, which summarize the temporal fluctuations in terms
of scaling exponents.  In Fig.\ 3, we exhibit by scatter plot
the values of $\alpha_1$ and $\alpha_2$ of all channels for
subject A.  The three points marked with circles correspond
to the channels shown in Fig.\ 2.  The error bars in Fig.\ 3
indicate the goodness of fit of the two regions by straight
lines.  Since the variability of the scaling exponents across
channels is large compared to the  error in fit, the different
values convey numerically meaningful information.  For this subject,
Region I exhibits better scaling behavior than Region II,  although
the error bars on $\alpha_2$ are not so large as  to call into
question the power-law description in Region II.

Overall, for subject A, the scaling exponents are in the  ranges:
$0.19<\alpha_1<1.44$ and $0.018<\alpha_2<0.489$.   Whereas
$\alpha_1$ is widely distributed, $\alpha_2$ is  sharply peaked
at 0.1 and has a long tail.   The value of $\alpha=0.5$
corresponds to random walk with no correlation among the various
time points. For $\alpha\neq 0.5$ there are correlations: Region
I corresponds to short-range correlation, Region II long-range,
with $\kappa$ giving a quantitative demarkation between the two.
In most channels  we find $\alpha_1 > \alpha_2$, although there
are a few  where $\alpha_1 \approx \alpha_2$.  The scatter plots
of all other  subjects are similar in general features to the one
shown  in Fig.\ 3, but vary in detail from subject to subject.
It is impractical to show them all in this paper.  Evidently,  it
is desirable to find a way to quantify succinctly these  128 pairs
of numbers so that one can effectively compare  the results
across subjects.

A scatter plot such as Fig.\ 3 reveals very well how the
$\alpha_i$ exponents of all the channels are related to one
another. However, it shows nothing about the locations of the
channels on the scalp. To show that, we use the topographical
plots of $\alpha_1$ and $\alpha_2$  separately, as in Fig.\ 4, to
exhibit the spatial structure of the signals extracted. The
dissociation of $\alpha_1$ from $\alpha_2$  is a price paid to
gain the spatial perspective on the scalp. Topographic plots
such as this may be useful for specifying the location
of focal features, e.g., associated with particular brain
functions and/or pathologies. Thus topographical and scatter plots
present different aspects of the fluctuation properties of the
brain electrical activity of any given subject. Both are
inefficient for comparison across subjects. What we need is a
global measure that describes the general, overall feature of all
$\alpha_i$ pairs in the form of a single parameter.

\section {Moments of the Scaling Exponents}

We propose to consider the moments of the scaling exponents. In
general, if we have $N$ numbers, $z_j, j=1,\cdots, N,$ we can
calculate the moments
\begin{equation}
G_q = {1\over N} \sum_{j=1}^N z_j^q , \label{5}
\end{equation}
where $q$ is a positive integer \cite{gar}. The information
contained in the first $N$ moments (i.e., $q=1,\cdots,N$) is
enough to reproduce all the $z_j$	by inversion. However, we may be
interested in only a few of the $G_q$ with lower order $q$, each of which
contains some information of all the $z_j$. In our
present problem we have
$N=128$, and we shall consider the first ten orders, $1\le q\le
10$. That is a significant reduction of numbers, a process worth
investigating.

Before calculating the moments of $\alpha_i$, let us see how
those values are distributed. Let $x$ be either $\alpha_1$ or
$\alpha_2$. Since no value of $\alpha_i$ has been found to exceed
1.5 in the subjects we have examined, we consider the interval
$0\leq x\leq 1.5$. Divide that interval into $M$ equal cells,
which for definiteness we take to be $M=150$ here. Let the cells
be labeled by $m=1,\cdots, M$, each having the size $\delta
x=1.5/M$. Denote the number of channels whose
$x$ values are in the $m$th cell by $n_m$. Define
\begin{equation}
 P_m = n_m\,/\,N.  \label{6}
\end{equation}
It is the fraction of channels whose $x$ values
are in the range $(m-1)\,\delta x \leq x < m\,\delta x$. By
definition, we have $\sum_{m=1}^M\,P_m=1$. In Fig.\,5 we show as
an illustration the two graphs of $P_m$ for subject A. The two
graphs correspond to $\alpha_1$ and $\alpha_2$, and are, in
essence, the projections of the scatter plot in Fig.\,3 onto the
$\alpha_1$ and $\alpha_2$ axes. From Fig.\,5 we see that
$\alpha_1$ is widely distributed, while $\alpha_2$ is not. $G_1$
gives the average, and $G_2$ is related to the width.

Since the fluctuation of $m$ in $P_m$ should be measured relative to
its mean, let us consider the normalized moments \cite{rch}
\begin{equation}
M_q^{(i)}=G_q^{(i)}\left/ \left(G_1^{(i)}\right)^q\right.=
\sum_{m=1}^M m^q P_m^{(i)} \left/
\left(\sum_{m=1}^M m P_m^{(i)}\right)^q\right. ,
\label{7}
\end{equation}
where $i=1$ or 2.
Since these moments are averages of $(m/\overline m)^q$, where
$\overline m$ is the average-$m$, they are not very sensitive to
$\overline m$ itself. They contain the essence of the
fluctuation properties of $\alpha_{1,2}$ in all channels.
In terms of the scaling exponents explicitly, let us
use $\alpha_i(j)$ to denote the  value of $\alpha_i$
for channel $j$ so that Eq.(\ref{7}) may be rewritten as
\begin{equation}
M_q^{(i)}={1\over N}\sum_{j=1}^N
\alpha_i(j)^q\, \left/ \left({1\over N}\sum_{j=1}^N
\alpha_i(j)\right)^q . \right.   \label{7.5}
\end{equation}

In principle, it is possible to examine also the moments for
$q<0$, which would reveal the properties of $P_m$ at low values
of $m$. However, the accuracy of our data is not too reliable
for low-$k$ analysis, since the 60 Hz noise due to ambient
electric and magnetic fields has not been cleanly filtered out.
In this paper, therefore, we restrict our study to only the
positive $q$ values.  For high $q$, the large $m/\overline m$
parts of $P_m^{(1,2)}$ dominate $M_q^{(1,2)}$.

In Fig.\,6 the $q$-dependences of $\ln M_q^{(1,2)}$ are shown
for the distributions exhibited in Fig.\,5 for
$2\leq q\leq 10$. They are approximately linear except for the
low values of $q$. The same type of dependencies on $q$ are found
for all subjects. In Fig.\,6 we show two straight lines that can
fit very well the nearly linear behaviors of ln
$M_q^{(i)}$ vs $q$ for $q \ge 5$. Thus for large $q$ we have
\begin{equation}
M_q^{(i)} \propto {\rm exp}\ (\mu_i\,q), \qquad q\ge 5.  \label{8}
\end{equation}
The linear extrapolations of the lines to lower values of $q$ show
the degree of deviation of the the calculated values of
$\ln M_q^{(1,2)}$ from linearity. Since ln$M_q^{(1)}$ and
ln$M_q^{(2)}$ behave so similarly in their departures from the
linear dependencies on $q$, we plot ln$M_q^{(2)}$ vs
ln$M_q^{(1)}$ in order to exhibit their direct relationship without
explicit dependence on $q$. We find that they are linearly related
over a wider range of values. This linearity is found to be true for
all subjects. The plots for three of them are illustrated in
Fig.\,7, where the straight lines are the linear fits. Thus the
implication is  that there exists a universal power-law behavior
\begin{equation} M_q^{(2)} \propto
\left(\,M_q^{(1)}\,\right)^{\eta}
\label{9}
\end{equation}
valid for all subjects examined. From Eqs.(\ref{8}) and (\ref{9}) we
obtain
\begin{equation}
\eta = \mu_2 / \mu_1  ,  \label{10}
\end{equation}
but now $\eta$ is meaningful for all $q$ (except for the lowest
points) and in that sense independent of $q$. Thus we have
discovered a global measure $\eta$ that characterizes all
$\alpha_i$ values of a subject, and varies from subject to
subject. We postpone the display of all the
$\eta$ values for all subjects until later.

To understand the exponential behavior in Eq.(\ref{8}), we note
that $G_q$ is dominated by large $z_j$ when $q$ is large, as is
self-evident in Eq.(\ref{5}). For asymptotically large $q$, we
have $G_q\propto {\rm exp}\ (q \,{\rm ln}\,z_{\rm max})$, where
$z_{\rm max}={\rm max}\,\{z_j\}$. For intermediate $q$, all large
values of $z_j$ can make important contributions, and the
exponential dependence on $q$ can still prevail. The denominator
in Eq.(\ref{7}) is $G_1^q=\bar z^q={\rm exp}\ (\,q\,{\rm ln} \bar
z\,)$,  where $\bar z$ is the average of $z_j$, so it is also
exponential for any $q$. It is therefore clear that Eq.(\ref{8})
follows, and that $\mu_i$ depends on all $z_j$ with more weight on
the large $z_j$ values. The power-law behavior of Eq.(\ref{9})
implies that the exponent $\eta$ is independent of $q$ and that
all $\alpha_i$ values are relevant contributors to the universal
behavior. This is an important point  worth emphasizing: the
independence of $\eta$ on $q$ implies that the whole spectra of
$\alpha_1$ and $\alpha_2$ are summarized by the one index $\eta$.
 The fact that $\eta$ varies from subject to subject is a
consequence of the variability of all 128 pairs of
$(\alpha_1,\alpha_2)$ across the subjects, and offers the possibility
that $\eta$ can be used as a discriminating representation of the brain
state.

\section{Correlations of the Scaling Exponents}

The analysis in the preceding section treats the moments of
$\alpha_1$ and $\alpha_2$ separately. Only in the last step are the
global properties embodied in $M_q^{(1)}$ and $M_q^{(2)}$ related
through the exponent $\eta$ in Eq.(\ref{9}). In that approach the
pairing of
$\alpha_1$ with
$\alpha_2$ in each channel is not taken into account. However, we
know that there are channels, such as Ch.\ 1 in Figs.\ 1 and 2, where
the absence of a dominant mode of oscillation results in $\alpha_1
\approx \alpha_2$. Thus the correlation between the two scaling
exponents is an important feature that should be explored and
quantified. To that end we define
\begin{equation}
\beta=\alpha_2 / \alpha_1   \label{11}
\end{equation}
for each channel. In most cases we have $\beta<1$, but $\beta>1$
is possible and, by its rarity, noteworthy.

From a scatter plot, such as Fig.\,3, it is possible to visualize
the $\beta$ distribution, since $\beta$ is just the slope of a line
from the origin to each point. We show in Fig.\,8 the $\beta$
distributions for the same three subjects as those in Fig.\,7.
Subject B is chosen for display because it has the largest $\eta$,
while subject C is chosen because it has several
$\beta$ values that exceed 1.

To summarize the 128 values of $\beta_j$ for each subject, we
apply to them the moment analysis that is developed in Sec.\,IV.
Let us therefore define
\begin{equation}
N_q={1\over N} \sum_{j=1}^N \beta_j^q \left/ \left({1\over N}
\sum_{j=1}^N \beta_j\right)^q \right..  \label{12}
\end{equation}
The $q$ dependence of ln\,$N_q$ for the same three subjects are
shown in Fig.\,9. Again, linear fits are very good. Thus we have
\begin{equation}
N_q\propto {\rm exp} (\,\nu q\,)  \label{13}
\end{equation}
with a distinct $\nu$ for each subject. Clearly, the ones with
wide $\beta$ distributions  relative to their means have  higher
values of $\nu$.

We now have found two indices, $\eta$ and $\nu$, for each subject.
They describe different aspects of the scaling exponents. To
display those values, it is illuminating to show the scatter plot
of $(\eta,\nu)$, as in Fig.\,10, which has 18 points for the 18
subjects studied. The subjects A, B and C are denoted by distinct
symbols, same as in Fig.\,9. We see that the points in Fig.\,10
appear to form a band, roughly correlated in $\eta$ and
$\nu$. Only subject C has a value of $\nu$ that lies above the
band, and it is C who has several
$\beta$  values above 1. Whether that is an anomaly carrying some
physiological significance is an issue outside the scope of this
paper, since we assume no knowledge of the physical conditions of
the subjects. We are similarly ignorant at this point about the
meaning of the spread of the $(\eta,\nu)$ values. However, it is
encouraging that the  scatter plot in $(\eta,\nu)$ is widely
distributed for the 18 subjects, since it offers the possibility
of being a discriminating tool, quite different from the
alternative scenario, if the parameters determined in an
analysis had turned out to be nearly the same for all subjects.

\section{Conclusion}

Recognizing that the brain is a highly nonlinear system, we have
explored a possible way of analyzing the EEG time series that
avoids the assumptions made in linear analyses and in chaos
studies. By studying the fluctuations from linear trends defined
over varying time scale, we have found two scaling regions in
which the RMS fluctuations can be characterized by two
dimensionless scaling exponents,
$\alpha_1$ and $\alpha_2$, for each channel. We then performed
moment analyses to reduce the large number of pairs of
$(\alpha_1,\alpha_2)$ to simple and manageable forms. The two
types of independent moments,
$M_q^{(i)}$ and $N_q$, yield two indices, $\eta$ and $\nu$, which provide
concise signatures of the nonlinear
behavior of all channels of the EEG signals.

Our emphasis in this paper has been on the method of analysis
rather than on the physiological interpretation of the results. For
the latter task it is necessary to have not only more data, but
also detailed information on the physical conditions of the
subjects so as to have a reference frame to calibrate the indices
obtained. However, working with 18 subjects is sufficient to
demonstrate the effectiveness of the method, to show the
universality of the scaling behaviors, to reveal the range of
variability of the indices derived, and to offer the possibility of
a new way of understanding human brain activity.

\vspace{1cm}

\begin{center}
{\bf Acknowledgment}
\end{center}

We are grateful to Prof.~Don Tucker and Dr.~Phan Luu  for
supplying the EEG data for our analysis.  We have also benefited from
the computational assistance of Wei He. This work was supported, in
part,  by the U.\,S.\,Department of Energy under Grant No.\,
DE-FG03-96ER40972, and the National Institutes of Health under
Grant No.\,R44-NS-38829.

\newpage
%\begin{bibliography}

%\end{bibliography}

% figure 1
\begin{figure}[ht]
\center\epsfig{figure=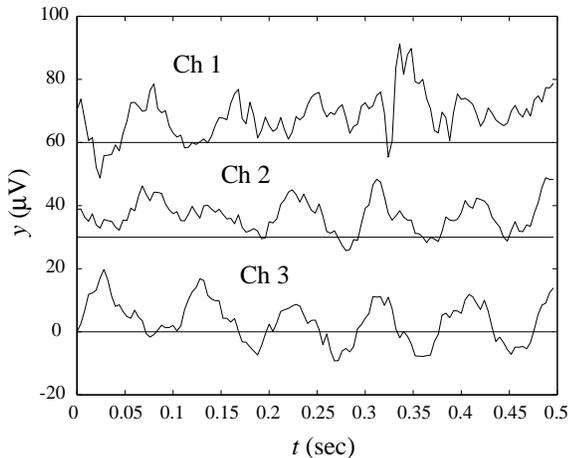,width=3.25in}
\caption{A sample of EEG time series in three channels. The
vertical scales of Ch.\ 1 and Ch.\ 2 are shifted upward by 60 and
30 $\mu$V, respectively.}
\end{figure}

% figure 2
\begin{figure}[ht]
\center \epsfig{figure=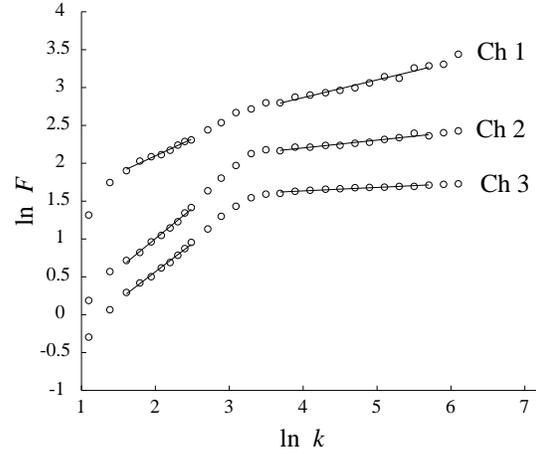,width=3.25in}
\caption{$F(k)$ vs $k$ for the three channels in Fig.\,1. The
vertical scales of Ch.\ 1 and Ch.\ 2 are shifted upwards by 1.0 and
0.5 units, respectively.}
\end{figure}

% figure 3
\begin{figure}[ht]
\center\epsfig{figure=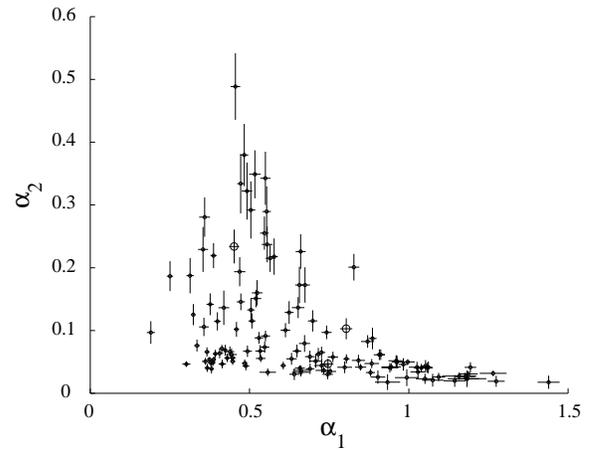,width=3.25in}
\caption{Scatter plot of $\alpha_2$ vs $\alpha_1$ for
subject A. The three channels exhibited in Figs.\,1 and 2 are
shown as circles.}
\end{figure}

% figure 4
\begin{figure}[ht]
\center\epsfig{figure=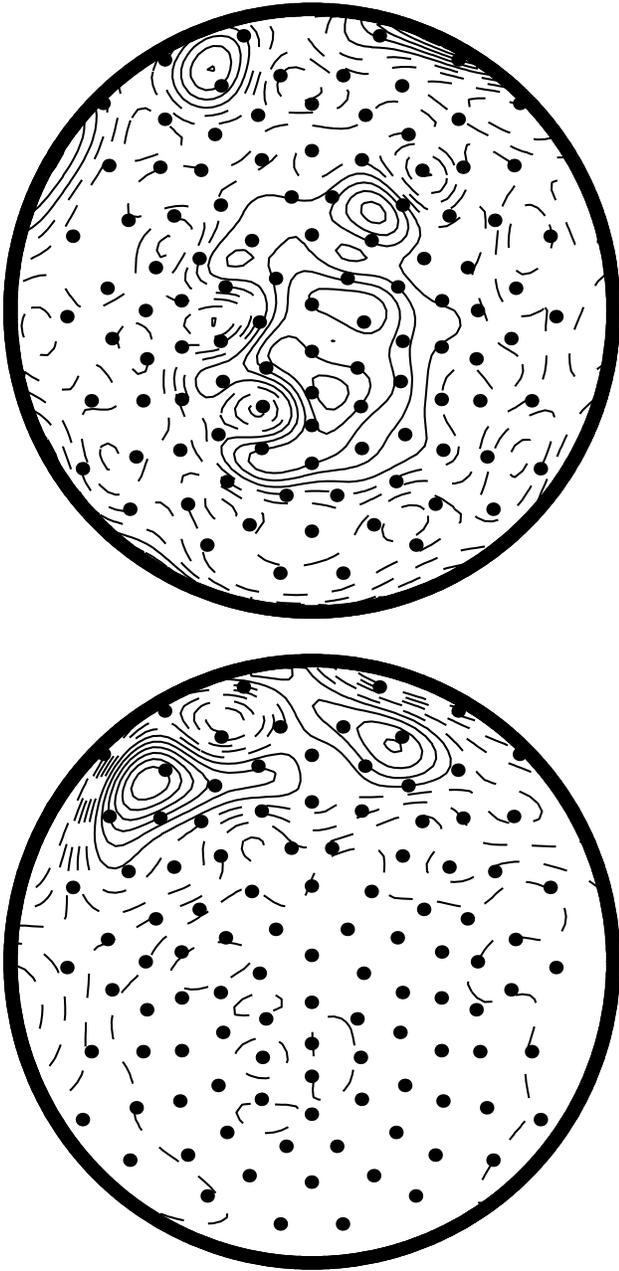,width=3.25in}
\caption{Topographical plots of $\alpha_1$ (top)
and $\alpha_2$ (bottom). In each figure, ten contour lines
are drawn within the data range: solid lines above the
mean, dotted lines below.}
\end{figure}

% figure 5
\begin{figure}[ht]
\center\epsfig{figure=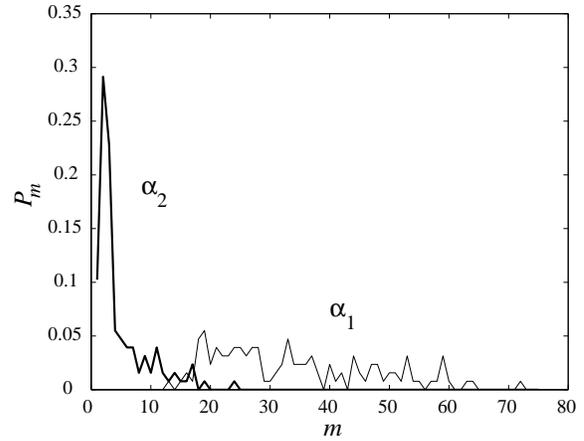,width=3.25in}
\caption{The distributions $P_m$ for $\alpha_1$ and
$\alpha_2$. The bin size in $\alpha$ for this plot is 0.02.}
\end{figure}

% figure 6
\begin{figure}[ht]
\center\epsfig{figure=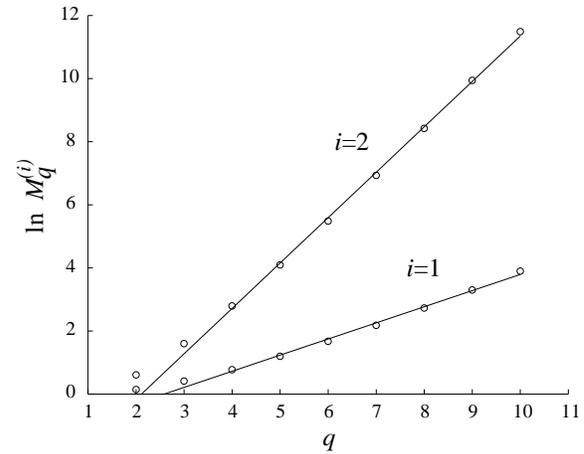,width=3.25in}
\caption{The $q$ dependence of $\ln M_q^{(i)}$ for
subject A. The straight lines are linear fits of the points for
$q \ge 5$.}
\end{figure}

% figure 7
\begin{figure}[ht]
\center\epsfig{figure=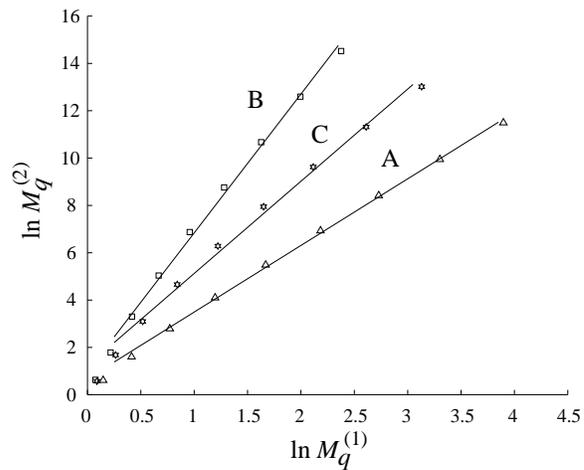,width=3.25in}
\caption{A log-log plot of $M_q^{(2)}$ vs $M_q^{(1)}$
for three subjects A, B, and C. The solid lines have the slopes
given by Eq.\,(\ref{10}).}
\end{figure}

% figure 8
\begin{figure}[ht]
\center\epsfig{figure=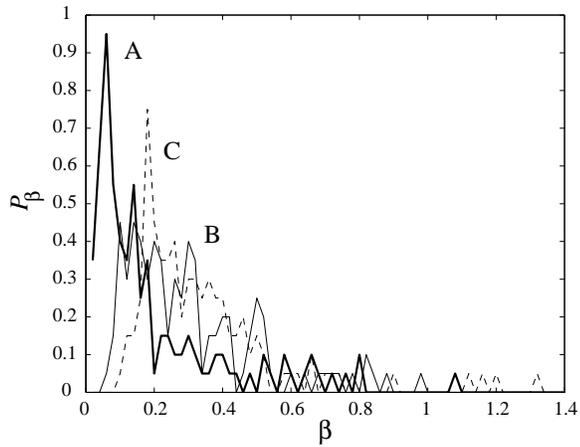,width=3.25in}
\caption{The distributions of the $\beta$ values of
the subjects A, B, and C.}
\end{figure}

% figure 9
\begin{figure}[ht]
\center\epsfig{figure=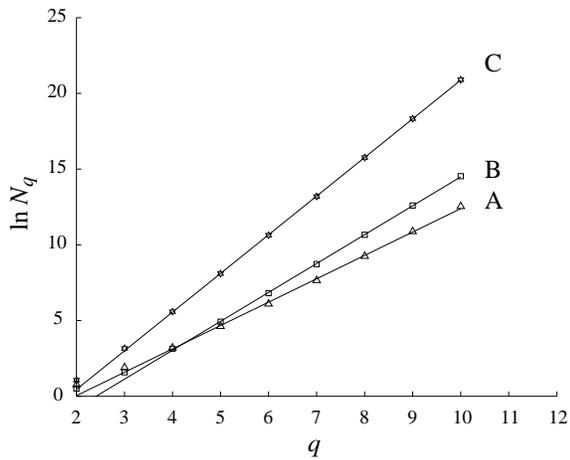,width=3.25in}
\caption{The $q$ dependence of $\ln N_q$ for
subjects A, B, and C.}
\end{figure}

% figure 10
\begin{figure}[ht]
\center\epsfig{figure=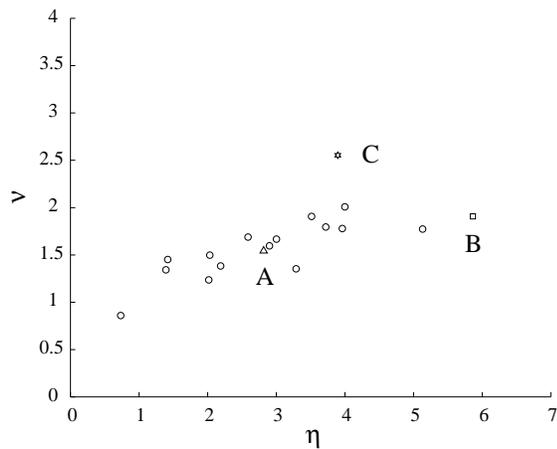,width=3.25in}
\caption{Scatter plot of $\nu$ vs $\eta$ for all 18
subjects, three of which have individual symbols: triangle (A),
square (B), and star (C).}
\end{figure}

\end{document}